\documentclass[3p,preprint,authoryear]{elsarticle}

\usepackage{amssymb}

\biboptions{comma}

\usepackage[figuresright]{rotating}

\begin{document}
\pagestyle{empty}

\begin{frontmatter}



 \title{Convenient liquidity measure for financial markets}

\author{Oleh Danyliv}
 \ead{oleh.danyliv@fidessa.com}
\author{Bruce Bland}
\author{Daniel Nicholass}
\address{Fidessa group plc, One Old Jewry, London, EC2R 8DN, United Kingdom} \tnotetext[t1]{Authors are grateful to
Steve Grob, Dr. Christian Voigt, Chris Moulang and Jeff Carey at Fidessa group plc for valuable discussions and
support.}
\tnotetext[t2] {The views expressed in this article are those of the authors and do not necessarily reflect the views
of Fidessa group plc, or any of its subsidiaries.}

\begin{abstract}
A liquidity measure based on consideration and price range is proposed. Initially defined for daily data, Liquidity
Index ($LIX$) can also be estimated via intraday data by using a time scaling mechanism. The link between $LIX$ and the
liquidity measure based on weighted average bid-ask spread is established.

Using this liquidity measure, an elementary liquidity algebra is possible: from the estimation of the execution cost,
the liquidity of a basket of instruments is obtained. A formula for the liquidity of an ETF, from the liquidity of its
constituencies  and the liquidity of ETF shares, is derived.
\end{abstract}


\end{frontmatter}

\pagestyle{headings} \setcounter{page}{1}

\section{Introduction}
\label{introcution}

Asset managers and ordinary investors care about liquidity, insofar as it affects the return on their investments,
because illiquid securities cost more to buy and sell. Illiquidity, which is opposite to liquidity, eats into an
investor's return. Another important aspect of liquidity is its effect on a portfolio evaluation: portfolio
liquidation, including illiquid assets, may reduce the value of a portfolio significantly. From another point of view,
a positive relationship between expected stock returns and illiquidity levels has been found \citep{amihud1986,
brennan1986, datar1998} which opens up new investment opportunities.

According to \cite{imf2002}, liquid markets should exhibit five characteristics: (i) {\it tightness}, which refers to
low transaction cost; (ii) {\it immediacy}, which represents the high speed of order execution; (iii) {\it depth},
which refers to the existence of limit orders; (iv) {\it breadth}, meaning small market impact of large orders; and (v)
{\it resiliency}, which means a flow of new orders to correct market imbalances.

It is no wonder, then, that with such wide requirements there is no single well-established liquidity measure.
Comprehensive reviews of different approaches to liquidity definition can be found in the following publications:
\cite{foucault2013, imf2002, goyenko2009}. We would like to mention the following groups of liquidity measures which
are widely used in theory and in practice.

{\bf Spread-related} measures capture the cost of the transaction. The absolute spread is the difference between the
lowest ask price and the highest bid price and is always positive. This type of liquidity has been widely investigated
recently \citep{chordia2001, hasbrouck1999, huberman2001}. Liquidity based on effective spread was studied in
\cite{barclay1999}. Roll's measure estimates liquidity from correlations in price change \citep{roll1984, george1991,
stoll2000}. The Glosten-Milgrom theory estimates the bid-ask spread on the basis of activity of informed traders
\citep{glosten1985}.

{\bf Volume-based} measures take into account the volume of the transaction. Average daily volume (ADV) is widely used
by practical traders to estimate how difficult is to trade a particular stock: the higher the ADV the more liquid the
stock and the easier it is to execute large orders without moving the market. \cite{amihud2002} illiquidity ratio,
$ILLIQ$, is one of the most widely used in the industry and is the daily ratio of absolute stock return to its dollar
volume averaged over some period. It is used by regulators to estimate liquidity trends. $ILLIQ$ illiquidity,
Hui-Heubel liquidity ratio and their relationship to the proposed liquidity measure are discussed in \ref{appendixA}.

{\bf Price-based} measures use variance of returns and test the resilience of markets to stock-related news. The market
efficiency coefficient described in \cite{bernstein1987} is a ratio of long-term return variance to the variance of
short-term returns.

In this paper we propose a liquidity measure which we have named Liquidity Index ($LIX$). This measure strongly
correlates and can be estimated by its instantaneous equivalent ($ILIX$) which is calculated using the order book data
and ADV. We show that the instantaneous liquidity measure has components related to market breadth, market depth,
market resilience and immediacy.

This paper proceeds as follows: Section \ref{section:LIX} introduces the liquidity measure used in this study. Section
\ref{section:intraday} presents the scaling mechanism which allows to use intraday data for liquidity calculation.
Section \ref{section:iLIX} introduces a related spread-based measure of liquidity. Sections \ref{section:basket} and
\ref{section:etf} provide formulae for combining the liquidities of individual instruments to calculate the liquidity
of a basket of securities and an ETF. Section \ref{section:conclusions} offers concluding remarks.


\section{The definition of $LIX$}
\label{section:LIX}

From a trader's perspective, the definition of liquidity could be formulated in the form of an answer to the following
question: {\em What amount of money can one invest without moving the market?} This is difficult to answer as it
depends on which strategy the trader uses to fill the order, how much time the trader has to execute it, how big the
order is compared to ADV, etc. However, refining the above question into {\em What amount of money is  needed to create
a daily single unit price fluctuation of the stock?} leads to the following, more precise definition of liquidity:

\begin {equation}
    Liquidity \sim
    \frac{Consideration}{Price \;Range} \equiv
    \frac{Volume \times Price} {High-Low}.
    \label{liquidity}
\end {equation}
The liquidity measure calculated in the above formula would range from thousands to billions, so to reduce this range
into more manageable numbers it is logical to take a logarithm of this amount. It is possible to do so because the
value calculated in (\ref{liquidity}) does not have units of measurement:
\begin {equation}
    LIX= \log_{10}
        \left ( \frac{V_{T} P_{Close}}{P_{High,T}-P_{Low,T}} \right ).
    \label{LIX}
\end {equation}
A logarithm with the base of 10 would create an index with a range from 5 to around 10, and would result in a simple
meaning: for US stocks the amount of capital needed to create \$1 price fluctuations could be estimated as $10^{LIX}$
of US dollars.

Liquidity measure (\ref{LIX}) in non-logged version (\ref{liquidity}) has a very simple form and we have found examples
of its use in the published literature. It resembles the \cite{amihud2002} illiquidity measure, which is a good proxy
for the theoretically founded Kyle's price impact coefficient $\lambda$ \citep{kyle1985}. In the UK government's
Foresight report \citep{linton2012} the measure (\ref{liquidity}) was used to investigate the evolution of market
liquidity with the following remark: ``in the current environment a plausible alternative to close to close return is
to use intraday high minus low return, since there can be a great deal of intraday movement in the price that ends in
no change at the end of the day".

\begin{figure}[htp]
    \centering
    \includegraphics[width=15cm]{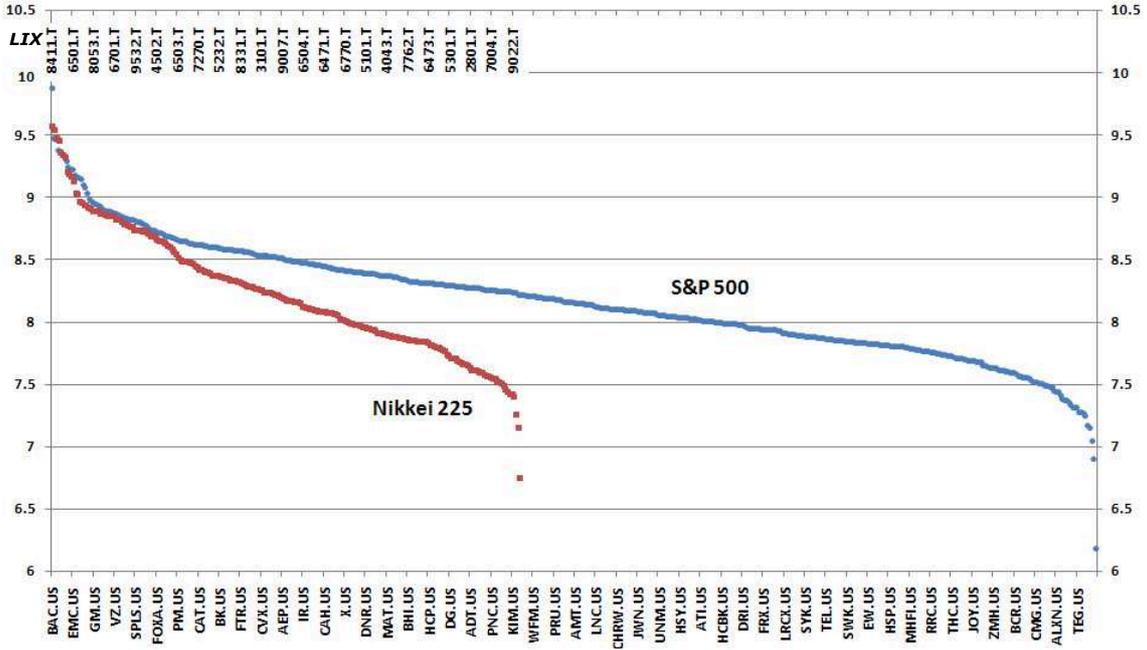}
        \caption{Shares of S\&P 500 and Nikkei 225 sorted by the liquidity measure calculated on 20 Nov 2013.}
    \label{figure:lix_sp500_nikkei}
\end{figure}

From our perspective, while the definition (\ref{LIX}) is similar to the Amihud illiquidity measure and to the
Hui-Heubel Liqidity Ratio (see \ref{appendixA}), there are a number of distinct differences. First, it uses already
mentioned ``high minus low" instead of daily return values; second, it is defined on daily data and does not use
averaging. Taking a logarithm introduces an important smoothing of this measure. We have found that the distribution of
logarithms of the daily volume (the main contributor to the liquidity value) could be modelled with good accuracy by a
Gaussian distribution - a calculation of the average of $LIX$ with a constant share price would correspond to a
geometrical average rather than standard arithmetic average.

The liquidity measure (\ref{LIX}) has the following advantages: it eliminates the currency value from calculations and
instruments on different international markets can be compared directly, and it is extremely easy to calculate because
all the information required can be obtained from a newspaper or a free internet site. Figure
\ref{figure:lix_sp500_nikkei} shows the constituent stocks of the S\&P 500 and Nikkei 225 indices sorted by the
liquidity measure. The liquidity decays almost linearly from the most liquid to illiquid stocks and has a convenient
range, from value five for the illiquid stocks to around ten for the most liquid.

\section{The intraday liquidity measure and liquidity time scaling}
\label{section:intraday}

Liquidity measure (\ref{LIX}) requires knowledge of the price range during the course of the day and is only available
after the trading session has closed. To calculate liquidity today we ought to use data and price action from the last
trading day, but this is neither convenient nor relevant because today's trading session could be more volatile. That
is why it is important to have a proxy for the liquidity measure using current data which is available from the start
of the session. Changing the underlying time interval in the liquidity measure will produce results that will not be
comparable with the measure defined in Section \ref{section:LIX}. To derive an estimate for the necessary scaling, let
us define intraday liquidity over the measured time $t$ in a similar manner to (\ref{LIX}):

\begin {equation}
    LIX_{t}= \log_{10}
        \left ( \frac{V_{t} P_{t}}{P_{High,t}-P_{Low,t}} \right ).
    \label{LIXt0}
\end {equation}
To understand how {\it LIX} should be scaled with time, we have to investigate the time scaling of its components on an
individual basis.

{\bf Price.} The price of a stock paying no dividends grows according to the level of the interest rate and is
negligibly small on intraday time intervals. The price at the end of the period $P_{t}$ is the most accurate valuation
of the stock at the time and this value should be used in the formula. Because of the log, price selection will not
have significant impact on the liquidity value.

{\bf Volume.} In general, the traded volume is a non-linear function of time and should be described by a volume
profile which is a proxy of how the volume will behave on average during the course of the day. Usually it has a U
shape and is higher at the beginning and the end of the session. Inconveniently, the volume is seriously affected by
news. Let us ignore these effects and assume that the volume traded is a linear function of time:
\begin {equation}
    V_{t}= V_{T} \times (t/T).
    \label{volume}
\end {equation}

{\bf Price range.} The variance of returns growth is a linear function of time. The volatility (square root of
variance) increases proportionally to $t^{\frac{1}{2}}$. The price range during time $t$, $\Delta S_{t} = P_{High,t} -
P_{Low,t}$ is different from the volatility and in the general case it should be scaled as $t^{\alpha}$:
\begin{equation}
        \Delta S_{t}= \Delta S_{T} \times (t/T )^{\alpha}.
    \label{range}
\end{equation}
Using Monte-Carlo simulation, it is easy to check that for a random walk the scaling factor for the price range has
value $\alpha=\frac{1}{2}$. But it possible to include a different scaling factor (for example $\alpha = 0.6$) to take
into account the fat tails of real distributions.

Substituting formulae (\ref{volume}) and (\ref{range}) into the equation for the intraday liquidity, one can obtain
\begin {equation}
    LIX= \log_{10}
        \left ( \frac{V_{t} P_{t}}{P_{High,t}-P_{Low,t}} \right )
        + (1-\alpha)  \log_{10} ({T}/{t}),
    \label{LIXt}
\end {equation}
or, taking theoretical value $\frac{1}{2}$ for $\alpha$, the estimation for liquidity using an intraday liquidity can
be done using

\begin{equation}
    LIX= LIX_{t}+\frac{1}{2} \log_{10} (T/t).
\end{equation}

\begin{figure}[htp]
    \centering
    \includegraphics[width=12cm]{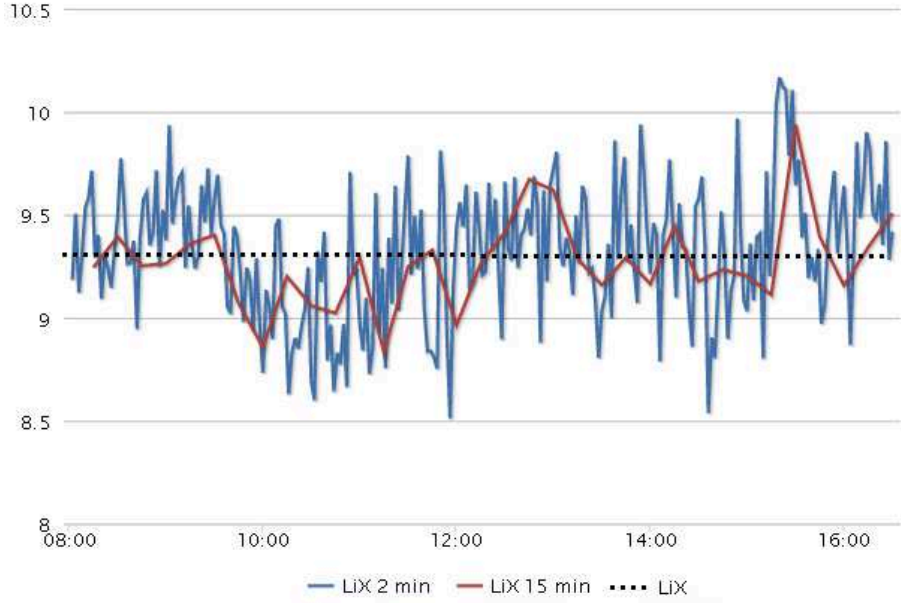}
        \caption{Intraday $LIX$ for London Stock Exchange traded stock BARC.L on trading session 01 Nov 2013 with theoretical
        scaling ($\alpha$=0.5) and using different time intervals to calculate the liquidity.
        The dotted line corresponds to the value of liquidity  $LIX = 9.3066$ calculated using definition (\ref{LIX}).}
    \label{figure:intraday}
\end{figure}

Figure~\ref{figure:intraday} shows that the intraday estimation for the daily liquidity is predominantly stable. The
estimations fluctuate around daily based value (\ref{LIX}) which is represented by a dotted line. Large changes in
intraday liquidity values are possible and they could be driven by announcements and news. Because the UK and US
markets are
 tightly linked, the opening of the US market was found to create a change in the liquidity value of UK stocks. A jump in
liquidity around 15:00 hrs is related to the ISM manufacturing index announcement on that day.

\section{Instantaneous liquidity measure}
\label{section:iLIX}

A problem that often occurs when we try to mathematically model real world events is that we come across the strange
reality of time, and its effect on what we are measuring. A good example of this is the concept of average speed. When
travelling between home and work each day, it takes a man approximately an hour. The distance between the two places is
thirty miles, so the man travels at an average speed of thirty miles per hour. The model works well, but it does not
tell us much about how he gets there or even if he spends time stationary waiting for a train! It is possible to
measure his speed at several points along the way, say on the train, and take a relatively precise measurement of his
speed. Say it was forty-five miles an hour. While this does not tell us when he will get to the office, it does give us
an approximate time. So if the train is slower, it might be reasonable to assume that it will take longer to get there.
If a large number of train speed readings are taken, an average could be used, and it is likely that a relationship
could be established between the journey time and speed at a single point in time.

The same principle applies when looking at an equity market during an active market session. While liquidity may appear
to be present and constant, it is in fact changing throughout the day, resulting in an average liquidity that can be
measured at the end of the day. Like the man's journey home, a completely different measure can be taken of the
instantaneous liquidity present, from the average end of day liquidity value. The two measurements should, however, be
well correlated to prove that they approximate for the same underlying value, like the speed of the train and total
journey time. Let us define the instantaneous liquidity using the information which is available at any time and which
is not related to historical data:
\begin {equation}
    LIXI_{\tau}= \log_{10}
        \left (
            \frac{ (V_{Bid}+V_{Ask}) P_{Mid} } {\overline{P_{Ask}} - \overline{P_{Bid}} }
        \right ),
    \label{LIXI0}
\end {equation}
where the midprice of an instrument $P_{Mid}=(P_{Ask, 1} + P_{Bid, 1})/2$ is an average between the best bid and ask
prices, $P_{Ask, i}$ is the ask price on the $i$-th level starting from the touch price, $V_{Bid}$ is the total bid,
and $V_{Ask}$ is the total ask volume in the limit order book calculated over available $N$ levels of the market depth:
\begin{eqnarray}
  V_{Ask}= \sum_{i=1}^{N}  V_{Ask,i}, \quad
  V_{Bid}= \sum_{i=1}^{N}  V_{Bid,i}.
\end{eqnarray}
$\overline{P_{Bid}}$ and $\overline{P_{Ask}}$ are the volume weighted averages of the bid and ask prices in the limit
order book:
\begin{eqnarray}
   \overline{P_{Ask}}= \frac {\sum_{i=1}^N P_{Ask,i}  V_{Ask,i}} {\sum_{i=1}^{N}  V_{Ask,i}}, \quad
   \overline{P_{Bid}}= \frac {\sum_{i=1}^N P_{Bid,i}  V_{Bid,i}} {\sum_{i=1}^{N}  V_{Ask,i}}.
\end{eqnarray}
The difference $\overline{P_{Ask}} - \overline{P_{Bid}}$ would be proportional to an effective spread for large market
order trades, big enough to wipe out all trades in the order book.

The liquidity measure (\ref{LIXI0}) has index ${\tau}$ to emphasise that in this form it is not directly comparable to
$LIX$. To find the link between the two different definitions of liquidity (\ref{LIX}) and (\ref{LIXI0}), let us
consider the volume traded. The liquidity formula (\ref{LIX}) is based on a whole day observation when, on average, the
ADV of shares is traded. Formula (\ref{LIXI0}) suggests that the equivalent of volume $V_{Bid} + V_{Ask}$ was traded on
the market. The market expects this volume to be traded during time:
\begin{equation}
 \tau = T \left ( \frac {V_{Bid} + V_{Ask}} {ADV} \right ),
\end{equation}
where $T$, as previously, is the length of the trading session. Using time $\tau$ and time scaling defined in Section
\ref{section:intraday}, we can correct equation (\ref{LIXI0}) to make it comparable to the day's liquidity value:
\begin {eqnarray}
    LIXI = \log_{10}
        \left (
            \frac{ (V_{Bid}+V_{Ask}) P_{Mid} } {\overline{P_{Ask}} - \overline{P_{Bid}} }
        \right ) +
        (1 - \alpha)
        \log_{10}
        \left (
            \frac {ADV} { V_{Bid}+V_{Ask} }
        \right ),
    \label{LIXI}
\end {eqnarray}
where $\alpha$ is a scaling factor for a price range.

\begin{figure}[htp]
    \centering
    \includegraphics[width=12cm]{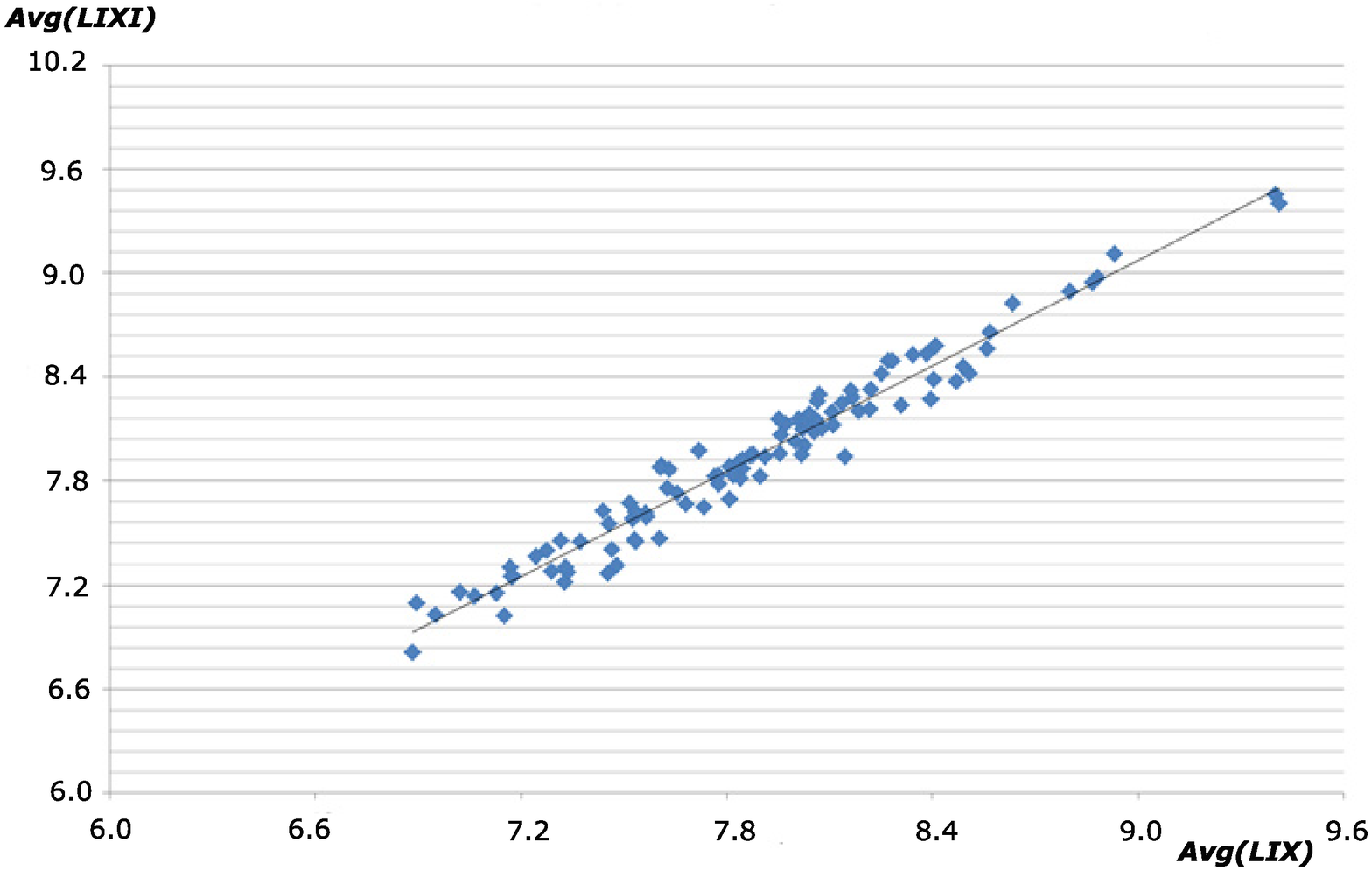}
        \caption{One day average $LIXI$ calculated with $\alpha = \frac{1}{2}$ vs one month average $LIX$ for all FTSE 100 stocks on 28 Jan 2013.
        The regression line has equation $y = 1.0147x - 0.0199$ with
        coefficient of determination $R^2 = 0.9596$.}
    \label{figure:LIXI}
\end{figure}

Figure~\ref{figure:LIXI} shows how strongly the instant liquidity $LIXI$ and the original liquidity $LIX$ are linked to
each other. The $R^2$ of the regression line is close to unity at $0.9596$, indicating strong correlation. This result
is important because it shows that volume-based liquidity measure (\ref{LIX}) is tightly related and can be substituted
by a spread based measure (\ref{LIXI}).

The quoted spread $P_{Ask,1} - P_{Bid,1}$ is a measure of trading cost for small orders which can be entirely filled at
the best quotes. Large market orders will be partially filled at best bid and offer price and then ``walk down the
book", penetrating the market depth. Ignoring non-displayed liquidity \citep{bacidore2002}, the average execution price
for a buy market order of size $q$ would be $\bar{a}(q)$, and the average execution price for a sell market order of
the same size would be $\bar{b}(q)$. Following definitions given by \cite{foucault2013}, the weighted average bid-ask
spread for an order of size $q$ is thus $S(q) = \bar{a}(q) - \bar{b}(q)$. The spread for an order big enough to wipe
out the limit order book is $\overline{P_{Ask}} - \overline{P_{Bid}}$. Defining relative spread as:
\begin{equation}
 s = \frac {\overline{P_{Ask}} - \overline{P_{Bid}}} {P_{Mid}},
\end{equation}
after simple transformation, and taking theoretical value $\alpha = \frac{1}{2}$ for scaling, the formula for instant
liquidity would have a simple form:
\begin{equation}
LIXI = -\log_{10} (s)
        + \frac{1}{2} \log_{10}
        \left (
            V_{Bid}+V_{Ask}
        \right )
        + \frac{1}{2} \log_{10}
        \left (
            ADV
        \right ).
    \label{LIXI1}
\end{equation}
This formula explicitly shows the intrinsic structure of liquidity measure (\ref{LIXI}). First of all it includes the
{\it transaction cost liquidity}, which is responsible for the {\it tightness} of markets as it is inversely
proportional to the average spread (negative logarithm dependency). A spread value is the most commonly used definition
of illiquidity. The second term is a {\it market depth liquidity} which has a simple intuitive interpretation: if the
stock has a lot of order volume in the order book, it is liquid.

The third term in (\ref{LIXI1}) is proportional to ADV, which is a raw measure of liquidity for practical traders. If
the daily volume traded in an instrument is divided by the total number of trades for the day, we can better understand
how the volume trades throughout the day. Effectively a liquid stock will trade more frequently than an illiquid one.
This frequency of trading is important as it gives an estimation as to the likely replacement frequency also. The ADV
therefore could be presented as a product of an average volume size, $ \overline{Size}$, and an average number or
trades, $\overline{N_{Trades}}$, resulting in:
\[
    \log_{10} \left ( ADV \right ) \sim \log_{10} \left ( \overline{Size} \right )
    + \log_{10} \left ( \overline{N_{Trades}} \right ).
\]
Market {\it breadth} is often described as the small market impact of big orders. Liquidity measure $LIXI$ implicitly
includes an average trade size which, if large, allows large orders to be matched on the market more easily, thereby
giving additional {\it breadth} to the market. The dependency of the liquidity measure on an average number of trades
implicitly describes the {\it resiliency} of the market (a flow of new orders), as well as the market {\it immediacy}:
the number of trades is inversely proportional to the execution time of a single order.

\section{Liquidity of a basket}
\label{section:basket}

It is standard practice for an informed market participant working a large order to use an execution algorithm, such as
VWAP (which tracks volume weighted average price) or Percentage of Volume (PoV). An order filled using such an
algorithm will tend to accumulate both passive and aggressive trades resulting in fills at bid and ask prices (see, for
example, \cite{leshik2011}). In general, the execution price $P'$ is the volume weighted average price. If the trading
activity generated a price move $\Delta P$, then the market participant using the algorithms will purchase shares
during the full period of the price rise and the average execution price could be estimated as $P' = P + \frac{\Delta
P}{2}$. After the trade, in resilient markets, the price will return to its previous level $P$. The transaction cost
$C_n$ of purchasing $n$ shares of stock is:
\begin{equation}
C_n = (P' - P) \times n = \frac{\Delta P} {2} n .
\end{equation}
The price range generated by trading $n$ shares during time $t$ could be estimated from  (\ref{LIXt}) as:
\begin{equation}
    \Delta P(n) = \frac{nP}{10^{LIX}} \left(\frac{T}{t}\right)^{1-\alpha},
\label{tran_cost}
\end{equation}
and the transaction cost of purchasing $n$ shares of a stock is:
\begin{equation}
C_n^{max} = \frac{1}{2} \frac{n^2 P} {10^{LIX}}  \left(\frac{T}{t}\right)^{1-\alpha}.
\end{equation}
The cost is a quadratic function of the number of traded shares. This is the estimation of the maximum cost when only
one investor is active on the market and he or she is trying to purchase all shares during a short period of time $t$.
In practice, to minimise market impact, a trader would split the order into slices and, after each trade, would allow
the market to recover from the impact of the order. There is a trade-off between the cost associated with moving the
price and the cost associated with the exchange fees. In an ideal case, where there is no exchange fee, the best
strategy would be to trade only one share at a time $t$ and wait for the market to recover. Then the trader will create
$n$ slices and the transaction cost reduces to:
\begin{equation}
C_n = n \times C_1 = \frac{1}{2} \frac{n P} {10^{LIX}}  \left(\frac{T}{t}\right)^{1-\alpha}. \label{cost_stock_n}
\end{equation}
It is useful to introduce the transaction cost of a single currency unit of investment ($\$1$ for US stocks or $\pounds
1$ for UK stocks) by dividing this value by the total market value of $n$ shares ($n P$):
\begin{equation}
{\bf C} = \frac{C_n}{n P} = \frac{1} {10^{LIX}}  \left \{  \frac{1}{2} \left(\frac{T}{t}\right)^{1-\alpha} \right \}.
\label{cost_stock}
\end{equation}
The first term in this expression is the inverse liquidity and the expression in curly brackets corresponds to the
efficiency of the trader and what time interval he or she had chosen to execute the order.

Let us consider a basket with current market value $M$ which consists of $N$ different stocks. Each stock is
represented by $n_i$ amount of shares or by $m_i = n_i \times P_i$ amount of money invested:

\begin{equation}
M = \sum_{i=1}^{N} m_i = \sum_{i=1}^{N} n_i P_i.
\end{equation}
In terms of the money, each price unit of investment is split into fractions $\beta_i = \frac{m_i}{M}$. The transaction
cost of acquiring this basket using the strategy of minimum market impact (\ref{cost_stock_n}) is:
\begin{equation}
C_{Bskt} = \sum_{i=1}^{N} C_{n_i} = \sum_{i=1}^{N} \frac{1}{2} \frac{m_i} {10^{LIX_i}}
\left(\frac{T}{t}\right)^{1-\alpha}.
\end{equation}
Each price unit  of investment will generate the transaction cost:
\begin{equation}
{\bf C}_{Bskt} = \frac{C_{Bskt}}{M} = \sum_{i=1}^{N} C_{n_i} = \sum_{i=1}^{N} \frac{\beta_i} {10^{LIX_i}} \left \{
\frac{1}{2} \left(\frac{T}{t}\right)^{1-\alpha} \right \}. \label{cost_basket}
\end{equation}
Comparing this result with the transaction cost per currency unit of investment for a single stock (\ref{cost_stock}),
the liquidity of the basket could be defined as:
\begin{equation}
    \frac{1} {10^{LIX_{Bskt}}} = \sum_{i=1}^{N} \frac{\beta_i} {10^{LIX_{i}}}.
\label{basket1}
\end{equation}
The liquidity of the basket corresponds to the liquidity of an instrument, such that the transaction cost of one
currency unit of investment in this instrument and an investment into the basket are the same. For the $LIX$ of the
basket:
\begin{equation}
    LIX_{Bskt} = -\log_{10} \left ( \sum_{i=1}^{N} \frac{\beta_i} {10^{LIX_{i}}} \right ).
    \label{LIX_basket}
\end{equation}
This formula provides a general rule for how to combine liquidities of individual instruments and does not depend on
trading time or the value of the investment. \ref{appendixB} provides simple test examples for this formula.

\section{Liquidity of an ETF}
\label{section:etf}

An exchange-traded fund (ETF) tracks a basket of underlying instruments traded on a stock exchange or a derivatives
market. This basket is traded on a stock exchange as a new synthetic security. This complex nature of an ETF suggests
that a new approach for liquidity calculation is required. For example, let us consider the following case of an ETF
with only one very liquid underlying stock, Ford Motor Company shares. Since it is possible for market makers to
convert ETF shares into shares of underlying, it is logical that the new ETF should be as liquid as Ford shares. But
because this fund is new and rarely traded, a direct calculation of the ETF's liquidity using formula (\ref{LIX}) will
produce an inadequately low result. For this reason the liquidity of an ETF should incorporate its dual nature as a
basket and as a traded synthetic security.

In order to account for this extra liquidity we return to equation (\ref{liquidity}) and the basic non-logged
definition. If the instrument is traded on multiple venues then, assuming no arbitrage, its price and price range
should be the same. It is easy to see that the liquidity for an instrument which can be traded on multiple venues is:

\begin {equation}
    L_{1,2} \sim \frac{V_{1,T} P_{Close}}{P_{High,T} - P_{Low,T}} + \frac{V_{2,T} P_{Close}}{P_{High,T} - P_{Low,T}},
    \label{liquidity_sum}
\end {equation}
which can be simply re-written as:
\begin {equation}
L_{1,2} = L_1 + L_2.
\end {equation}
If we take $L_1$ to be the liquidity of the basket and $L_2$ to be the liquidity of the ETF as an instrument,
$L_{ETF}$, we can use equation (\ref{LIX_basket}) and the $LIX$ value to obtain the following:
\begin {equation}
    L_{Bskt,ETF} = \frac{1} {\sum_{i=1}^{N} \frac{\beta_i} {10^{LIX_{i}}}} + L_{ETF}.
\end {equation}
Finally this can be converted to the combined $LIX$ result for underlying and ETF by converting to a logarithmic
representation:

\begin {equation}
    LIX_{Bskt,ETF} = \log_{10} \left ( \frac{1} {\sum_{i=1}^{N} \frac{\beta_i} {10^{LIX_{i}}}} + 10^{LIX_{ETF}}
    \right ).
    \label{etf}
\end {equation}
It is easy to see that this formula solves the problem of ETF liquidity measurement mentioned at the beginning of the
section.

Formula (\ref{etf}) is extremely useful for the estimation of the liquidity of start-up ETFs, which are not intensively
traded, as well as for calculation of the liquidity of well-established ETFs which are liquid in their own right - for
example, SPDR S\&P 500 ETF Trust, which tracks the performance of the S\&P 500 index.

\section{Conclusions}
\label{section:conclusions}

A new measure for liquidity is proposed which allows the user to estimate the liquidity of different instruments,
regardless of exchange or the currency in which they are traded. The measure allows for the instantaneous calculation
of its value or a historic measurement over the day. These two forms of the metric have proved to be well correlated
with each other, and allow for real-time monitoring of markets or historical calculation of liquidity. A time scaling
approach has been proposed which allows for the value of liquidity to be obtained intraday. It has been shown that
these formulae can be used in the calculation of liqidity for portfolios of both stock and ETF instruments. As accurate
knowledge about an instrument's liquidity is fundamental to many financial applications, uses for this liquidity
measure range from portfolio construction to market abuse detection systems.

\appendix

\section{Widely used volume-based liquidity measures}
\label{appendixA}
\subsection {Hui-Heubel Liqidity Ratio}

The Hui-Heubel liquidity ratio, as it is referenced in \cite{imf2002}, is described by the formula:
\begin{equation}
L_{hh} = \frac {(P_{High}-P_{Low})/P_{Low})}{\$V/(M \times E(P))} \label{HH},
\end{equation}
where $P_{High}$ is the highest daily price over the last five days, $P_{Low}$ is the lowest daily price over the last
five days $\$V$ is the total dollar volume traded over the last five days, $M$ is the number of shares outstanding and
$E(P)$ is the average closing price of the instrument over a five day period. Noticing that the total dollar volume is
approximately the average price multiplied by the total volume $V$:
\begin{equation}
\$V= \sum_j P_j V_j \sim E(P) V, \label{dollarvalue}
\end{equation}
we can write:
\begin{equation}
1/L_{hh} \sim
\left (
   \frac{ V P_{Low} } {P_{High}-P_{Low}}
\right ) \left (
   \frac{1}{M}
\right ) .
\label{A3}
\end{equation}
Taking into account that the number of outstanding shares is a constant value, formula (\ref{A3}) is very similar to
the $LIX$ formula in non-logged form (\ref{liquidity}). Because formula (\ref{HH}) contains the number of shares
outstanding, in the case of a stock split the liquidity measure would have a significant drop. This makes the
Hui-Heubel formula less reliable than the simpler $LIX$.

\subsection {Amihud illiquidity measure}

\cite{amihud2002} illiquidity measure is given by the formula:
\begin{equation}
ILLIQ = \frac{1}{N} \sum_{i=1}^N ILLIQ_i = \frac{1}{N} \sum_{i=1}^N \frac{|r_i |}{\$V_i} ,
\end{equation}
where $r_i$ is the stock return on day $i$ and $\$V_i$ is the dollar volume on day $i$, whereas $N$ is the number of
days. Taking into account (\ref{dollarvalue}), one can write that the $i$-th contribution is:
\begin{equation}
 \frac{1} {ILLIQ_i} \sim \frac {V E(P)} {|r_i |},
\end{equation}
which is similar to $LIX$ in form (\ref{liquidity}) but, instead of a price range, Amihud used an absolute return. The
problem with this definition is obvious: if the market was very volatile during the day, but the closing price was the
same as the open price, then that day would not have any contribution to the calculation of liquidity.

\section {Liquidity of a basket, test cases}
\label{appendixB}
Expression (\ref{LIX_basket}) can be tested on the following extreme cases:
\subsection {A basket of one instrument}

In this case, using notations given in Section \ref{section:basket}, $N = 1$ and $\beta_1 = 1$. Using formula
(\ref{LIX_basket}):
\begin{equation}
    LIX_{Bskt} = -\log_{10} \left ( \sum_{i=1}^{1} \frac{1} {10^{LIX_{i}}} \right ) = LIX_1 .
\end{equation}
The important trivial conclusion: the liquidity of the basket is the liquidity of the instrument.

\subsection {A basket of two instruments with similar liquidity}
For such a basket, $N = 2$, $LIX_1= LIX_2$   and investment ratios $\beta_1=\beta$, $\beta_2=1-\beta$:
\begin{equation}
    LIX_{Bskt} = -\log_{10} \left (  \frac{\beta} {10^{LIX_{1}}} + \frac{1 - \beta} {10^{LIX_{1}}} \right ) = LIX_1 .
\end{equation}
If the basket consists of two instruments with the same liquidity, then the liquidity of the basket is the liquidity of
the instruments.

\subsection {A basket with one liquid and one very illiquid instrument}
Let us consider a basket of two instruments with the same investment ratios $N = 2$,  $\beta_1= \beta_2 = \frac{1}{2}$.
The liquidity of the first instrument is very small: $LIX_1 \ll LIX_2$. Then the liquidity of the basket is:
\begin{equation}
    LIX_{Bskt} = -\log_{10} \left (  \frac{0.5} {10^{LIX_{1}}} + \frac{0.5} {10^{LIX_{2}}} \right ) \geq
    -\log_{10} \left (  \frac{0.5} {10^{LIX_{1}}} \right ) > LIX_1 + 0.3 .
\end{equation}
The liquidity of the basket of two instruments, where one instrument is very illiquid, is higher than the liquidity of
the most illiquid instrument.


\bibliographystyle{elsarticle-num}

\end{document}